\documentstyle[12pt,epsfig]{article}
\def\baselinestretch{1.0}
\advance\textheight by \topskip
\begin{document}
\tolerance=100000
\thispagestyle{empty}
\setcounter{page}{0}

\def\jhep #1 #2 #3 {{\it J.~High.~Energy~Phys.} {\bf #1} (#2) #3}
\def\plb #1 #2 #3 {{\it Phys.~Lett.} {\bf B~#1} (#2) #3}
\def\npb #1 #2 #3 {{\it Nucl.~Phys.} {\bf B~#1} (#2) #3}
\def\epjc #1 #2 #3 {{\it Eur.~Phys.~J.} {\bf C~#1} (#2) #3}
\def\jpg #1 #2 #3 {{\it J.~Phys.} {\bf G~#1} (#2) #3}
\def\prd #1 #2 #3 {{\it Phys.~Rev.} {\bf D~#1} (#2) #3}
\def\prep #1 #2 #3 {{\it Phys.~Rep.} {\bf #1} (#2) #3}
\def\prl #1 #2 #3 {{\it Phys.~Rev.~Lett.} {\bf #1} (#2) #3}
\def\mpl #1 #2 #3 {{\it Mod.~Phys.~Lett.} {\bf #1} (#2) #3}
\def\rmp #1 #2 #3 {{\it Rev. Mod. Phys.} {\bf #1} (#2) #3}
\def\cpc #1 #2 #3 {{\it Comp. Phys. Commun.} {\bf #1} (#2) #3}
\def\sjnp #1 #2 #3 {{\it Sov. J. Nucl. Phys.} {\bf #1} (#2) #3}
\def\Ord{\lsim}
\def\lsim{\:\raisebox{-0.5ex}{$\stackrel{\textstyle<}{\sim}$}\:}
\def\OOrd{\gsim}
\def\gsim{\:\raisebox{-0.5ex}{$\stackrel{\textstyle>}{\sim}$}\:}

\begin{flushright}
{CERN-TH/2002-291\\
           IPPP/02/59\\
           DCPT/02/118\\
           November 2002}
\end{flushright}

\begin{center}
{\Large \bf
  Heavy charged Higgs boson production \\[0.25cm]
at next generation $\gamma\gamma$
  colliders}\\[1.5 cm]
{\large Stefano Moretti$^*$}\\[0.15 cm]
{\it Theory Division, CERN,}\\
{\it CH-1211 Geneva 23, Switzerland}\\[0.15 cm]
{\rm and}\\[0.15 cm]
{\it Institute for Particle Physics Phenomenology,}\\
{\it University of Durham, Durham DH1 3LE, UK}
\\[0.5cm]
{\large Shinya Kanemura$^*$}\\[0.15 cm]
{\it Theory Group, KEK, 1--1 Oho, Tsukuba,}\\
{\it Ibaraki 305--0801, Japan}

\end{center}
\vspace*{\fill}

\begin{abstract}
{\vskip0.25cm\noindent\small
We investigate the scope of all relevant production modes of charged
Higgs bosons in the MSSM, with mass larger than the one of the top quark,
at future Linear Colliders operating in $\gamma\gamma$ mode
at the TeV energy scale. Final states with one or two $H^\pm$ bosons are 
considered, as produced by both tree- and loop-level interactions.}
\end{abstract}
\vskip1.35cm
{$~~~$Keywords:
Higgs Physics, Supersymmetry, Linear Colliders, 
Photon-photon Interactions}
\vskip1.75cm
\hrule
\vskip0.75cm
\noindent
\hskip0.00cm{\small $^*$E-mails: stefano.moretti@cern.ch,
kanemu@post.kek.jp.}\\
\vspace*{\fill}
\newpage

\noindent
In the Minimal Supersymmetric Standard Model
(MSSM) it is not unnatural to assume that the typical mass 
of the Supersymmetric (SUSY) 
partners of ordinary matter is at the TeV scale or above --
well in line with current experimental bounds -- this rendering the 
Higgs sector a privileged probe to access physics beyond the SM. 
In this respect, it would be intriguing to detect charged Higgs
states (henceforth denoted by $H^\pm$), as in this case one would
unquestionably be in presence of some non-standard phenomena. In fact, even 
the discovery of a (light) neutral Higgs boson, would leave open questions as 
to whether it belongs to the SM or else the MSSM, since in the so-called 
`decoupling regime' of the latter (i.e., when a hierarchy exists among
the masses of the five Higgs states: 
$M_{H^0}\sim M_{A^0}\sim M_{H^\pm}\gg M_{h^0}$) the
fundamental properties of such a particle (quantum numbers, couplings, 
branching ratios, etc.) would be 
the same in both models\footnote{In practice, decoupling occurs
for $M_{H^0}$, $M_{A^0}$ and/or $M_{H^\pm}$ around and above 200 GeV.}.

Rumours of a possible evidence of light charged Higgs bosons being
produced at LEP2 \cite{HpmExcessLEP} have faded away. One is now
left with a model independent limit on $M_{H^\pm}$, of order $M_{W^\pm}$.
However, within the MSSM, the current lower bound on a light Higgs boson
state, of approximately 120 GeV (from LEP2), can be converted into a minimal
value for the charged Higgs boson mass, of order 140 GeV or so (at small 
values of $\tan\beta$, the ratio of the 
vacuum expectation values of the two Higgs doublet fields\footnote{Which, 
together with $M_{H^\pm}$, or $M_{A^0}$ (the mass of the pseudoscalar Higgs
boson), uniquely defines the MSSM Higgs 
sector at tree-level.}). 
In the mass interval $140$ GeV $\lsim M_{H^\pm}\lsim m_t$, 
charged Higgs bosons could well be found at 
Tevatron (Run 2) \cite{Run2}, which has 
already begun data taking at $\sqrt s_{p\bar p}=
2$ TeV \cite{ioemono} at FNAL. In contrast, if $M_{H^\pm}\gsim m_t$ 
(our definition of a `heavy' charged Higgs boson), one will necessarily
have to wait for the advent of the Large Hadron Collider (LHC, $\sqrt s_{pp}
=14$ TeV) at CERN. Even there though, because of the dependence of the
production cross section of charged Higgs bosons upon $\tan\beta$, there 
is no certainty that these particles will be 
accessible to the experiments. This happens if $\tan\beta$ is in the so-called 
`intermediate' regime, starting at around 6 or 7 for $M_{H^\pm}\sim m_t$
and encompassing
more and more parameter space as $M_{H^\pm}$ grows larger, no matter
the channels in which the charged Higgs boson decays to, as long as the
latter only include ordinary SM objects and neutral 
Higgs states \cite{reviewHpm}. Not coincidentally, over the same area 
of the $(M_{H^\pm},\tan\beta)$ 
parameter plane, there is no coverage through the neutral 
Higgs sector of the MSSM either.

Lowering the SUSY mass scale may induce new interactions among
neutral/charged Higgs boson states and sparticles, so that
the former may abundantly be produced in the decay of the latter 
(gluinos and squarks for example \cite{Asesh}) or, alternatively,
new Higgs decay channels into light SUSY particles may well open at 
profitable rate (e.g., into chargino-neutralino pairs \cite{Mike}). This 
unfortunately implies a proliferation of MSSM parameters rendering the 
phenomenological analysis very cumbersome. 

With the option of an $e^+e^-$ Linear Collider (LC) \cite{LCs}
being possibly available
within a few years of the beginning of the LHC, also operating in
$e^\pm\gamma$ and $\gamma\gamma$ modes (both at an energy scale similar to the
one of the primary electron-positron design, i.e., $\sqrt s_{e^+e^-}=500$ GeV 
to 1.5 TeV), with the photons being generated via Compton back-scattering
of laser light \cite{backscattering}, it is very instructive
to assess the potential of this kind of machine in complementing the
LHC in the quest for such elusive, yet crucial particles for understanding
the Higgs mechanism. Besides, the ability to polarise the incoming
particles, both electron\footnote{Some proposals also exist
for polarising positrons \cite{positron}.} and photon 
beams\footnote{See, e.g., Ref.~\cite{polphotons} for an example of 
new physics effects which can be probed by using polarised $\gamma$-beams
to produce $H^\pm$ Higgs states.}, 
is a definite advantage 
of future LCs with respect to the LHC.

Historically, with some exceptions,
it was mainly the pair production modes of charged Higgs boson 
states, i.e., $e^+e^-\to H^-H^+$, $e^\pm\gamma\to e^\pm H^-H^+$ and 
$\gamma\gamma\to H^-H^+$, that were considered in some detail
\cite{epem,egam,gamgam}. However,
the exploitation of these channels alone may clearly be insufficient to clarify
the real potential of future LCs in investigating the Higgs sector, 
especially considering that in the MSSM framework twice the heavy $H^\pm$ 
mass values may 
mean that the rest mass of $H^-H^+$ pairs is already comparable 
to the minimal energy foreseen for these machines. Needless to say, 
whenever $2M_{H^\pm}$ exceeds $\sqrt s_{e^+e^-}$, the double Higgs modes
just mentioned are altogether useless and one has to revert to the case
of singly produced $H^\pm$ bosons. 

The potential of LCs operating via $e^+e^-$ and $e^\pm\gamma$ scatterings
in detecting MSSM charged Higgs bosons with mass $M_{H^\pm}\gsim m_t$ produced 
in single modes (as well as their interplay with the pair production channels)
has already been assessed in Refs.~\cite{epemH,egamH}. Here, we perform
 a similar study in the context of $\gamma\gamma$ 
interactions\footnote{In the case of photon-photon collisions,
charged Higgs bosons can also be produced as virtual states, e.g.,
in the loop entering $\gamma\gamma\to$ Higgs processes. Such channels
can be used as a means to distinguish between various possible Higgs 
scenarios, e.g.: SM, MSSM and/or a general 
Two-Higgs Doublet Model (2HDM) \cite{2HDM}.}.
%
%
%
Alongside 
\begin{equation}\label{HH}
\gamma\gamma\to H^-H^+\qquad({\mathrm{pair~production}}),
\end{equation}
we have considered several channels where only one charged Higgs boson is
produced, namely
\begin{equation}\label{tnH}
\gamma\gamma\to\tau^-\bar\nu_\tau H^+
\qquad(\tau\nu{\mathrm{~associated~production}}),
\end{equation}
\begin{equation}\label{btH}
\gamma\gamma\to b\bar t H^+
\qquad(bt{\mathrm{~associated~production}}),
\end{equation}
\begin{equation}\label{H02WH}
\gamma\gamma\to h^0 W^- H^+
\qquad
(h^0{\mathrm{~associated~production}}),
\end{equation}
\begin{equation}\label{H01WH}
\gamma\gamma\to H^0 W^- H^+
\qquad
(H^0{\mathrm{~associated~production}}),
\end{equation}
\begin{equation}\label{H03WH}
\gamma\gamma\to A^0 W^- H^+
\qquad
(A^0{\mathrm{~associated~production}}),
\end{equation}
\begin{equation}\label{WH}
\gamma\gamma\to W^- H^+
\qquad
(W^\pm{\mathrm{~associated~production}}).
\end{equation}

\begin{figure}[!t]
  \begin{center}
  \hskip-5.0cm
  \epsfig{file=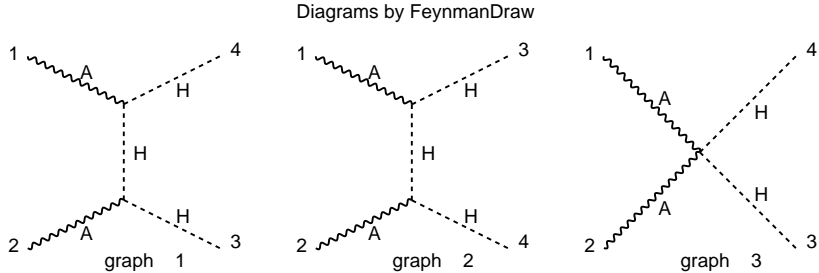,width=12cm,angle=0}\\
  \vspace{-16.5truecm}
  \caption{\small Feynman diagrams for process (\ref{HH}).
  The labels {\tt A and H} refer
  to a photon and a charged Higgs boson, respectively.}
  \label{feynman_graphs_HH}
  \end{center}
  \vspace*{-0.5cm}
  \end{figure}

\begin{figure}[!h]
  \begin{center}
  \hskip-5.0cm
  \epsfig{file=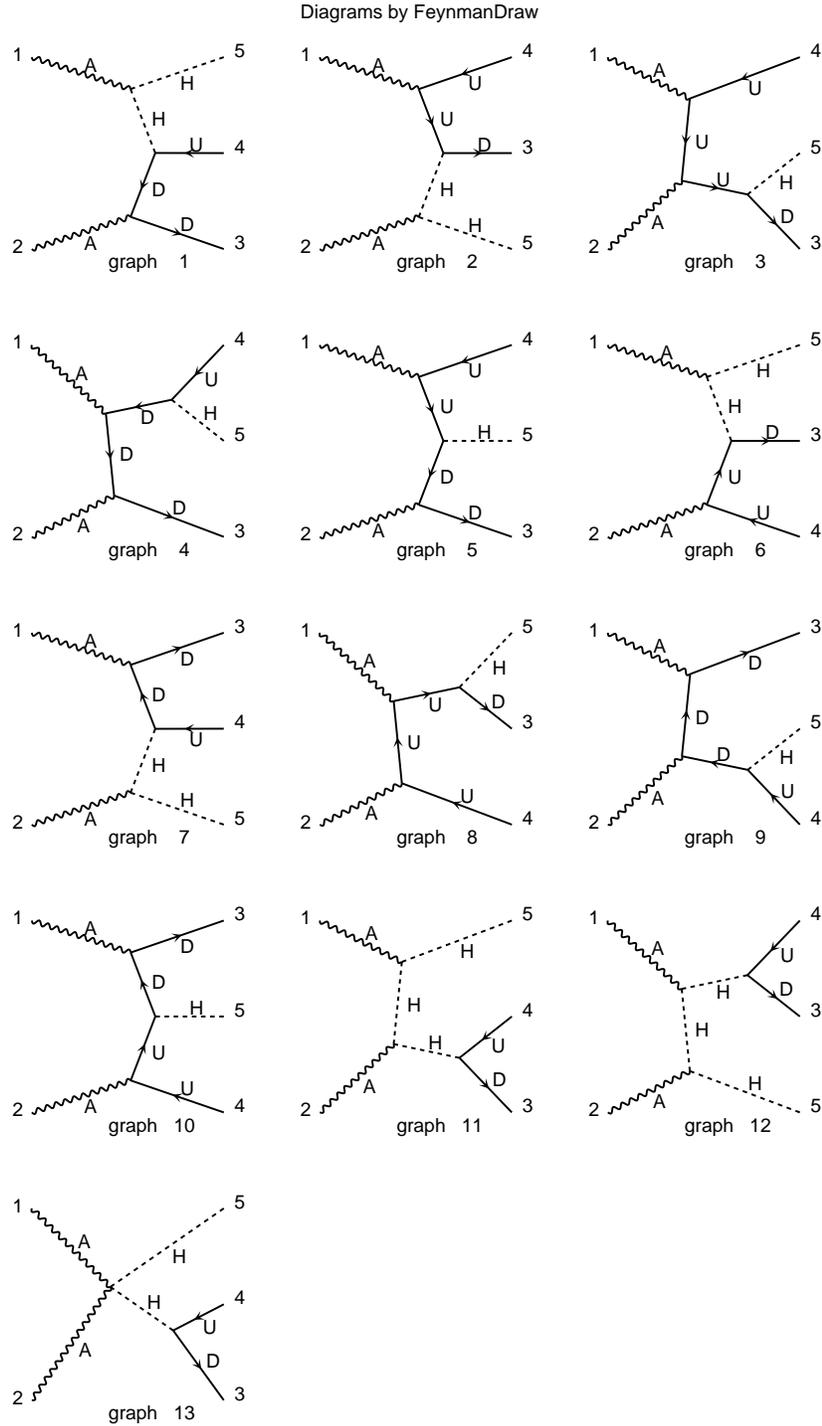,width=12cm,angle=0}\\
  \vspace{-1.25truecm}
  \caption{\small Feynman diagrams for processes of the type (\ref{tnH})--(\ref{btH}).
  The labels {\tt D/U, A and H} refer
  to a $d$/$u$-type (anti)quark, a photon and 
  a charged Higgs boson, respectively.}
  \label{feynman_graphs_DUH}
  \end{center}\end{figure}

\begin{figure}[!h]
  \begin{center}
  \hskip-5.0cm
  \epsfig{file=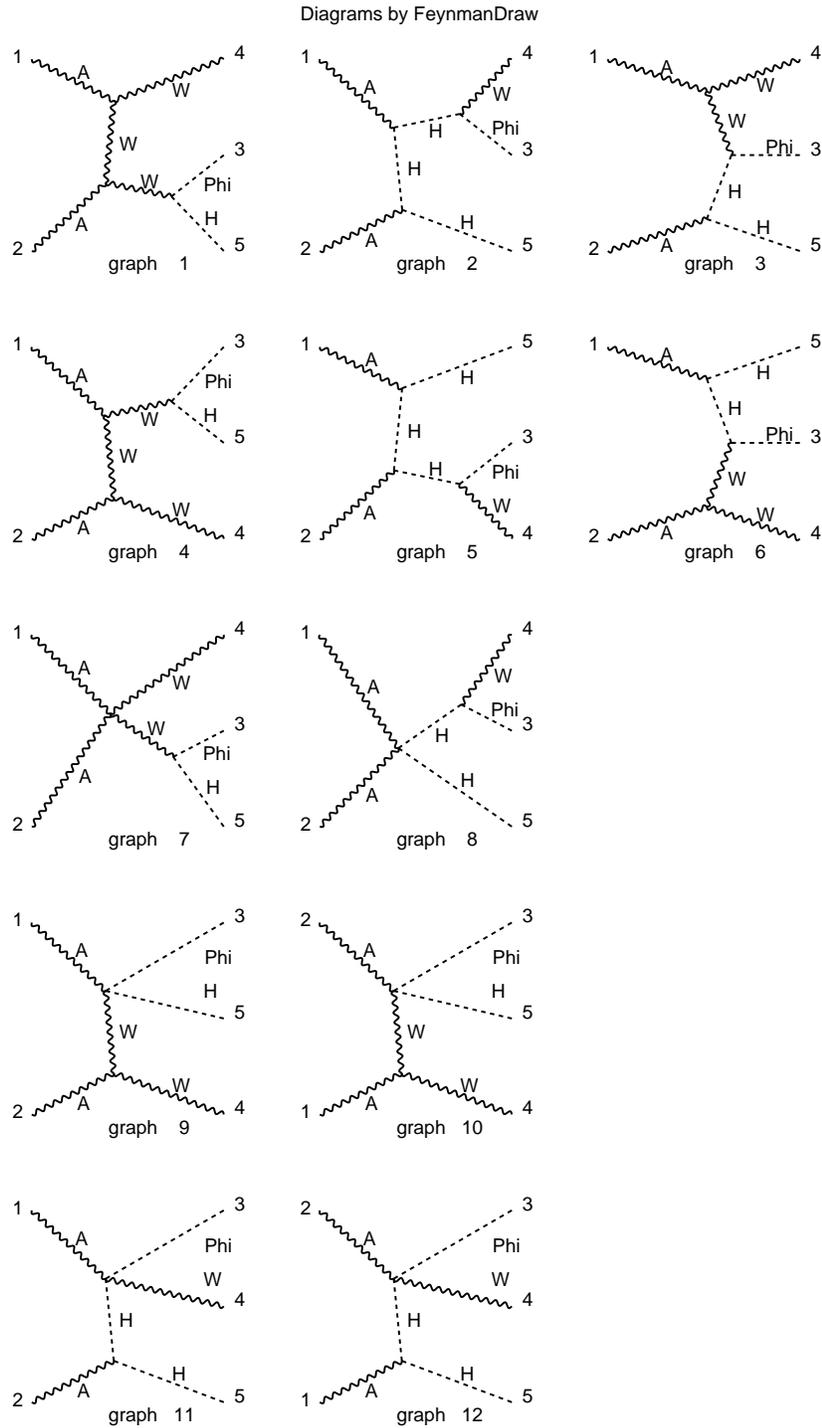,width=12cm,angle=0}\\
  \vspace{-1.25truecm}
  \caption{\small Feynman diagrams for processes of the type 
  (\ref{H02WH})--(\ref{H03WH}).
  The labels {\tt A, W and H(Phi)} refer
  to a photon, a $W^\pm$ gauge boson and 
  a charged(neutral) Higgs boson, respectively.}
  \label{feynman_graphs_H0WH}
  \end{center}\end{figure}

\begin{figure}[!t]
  \begin{center}
  \hskip-5.0cm
  \epsfig{file=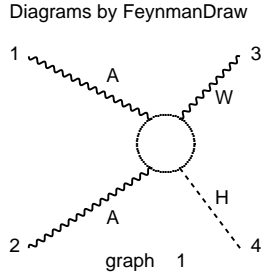,width=12cm,angle=0}\\
  \vspace{-16.5truecm}
  \caption{\small Feynman diagrams for process (\ref{WH}).
  The labels {\tt A, W and H} refer
  to a photon, a $W^\pm$ gauge boson and 
  a charged Higgs boson, respectively.
  The `blob' signifies all possible one-loop contributions,
  as seen in Figs.~1--2 of \protect{\cite{gamgamloop}}.}
  \label{feynman_graphs_WH}
  \vspace*{-0.5cm}
  \end{center}\end{figure}

The Feynman graphs associated to process (\ref{HH}) are found in
Fig.~\ref{feynman_graphs_HH}, those for reactions (\ref{tnH})--(\ref{btH})
in Fig.~\ref{feynman_graphs_DUH}, for (\ref{H01WH})--(\ref{H03WH})
see Fig.~\ref{feynman_graphs_H0WH}, whereas for
(\ref{WH}) refer to Fig.~\ref{feynman_graphs_WH}. 
All processes were calculated at leading order only. The first six
are tree-level processes whereas the last one originates at one-loop level.
All these we have computed by means of the
{\tt HELAS} libraries \cite{HELAS}, with the exception the one-loop channel 
(for which we have adapted the calculations
of Ref.~\cite{gamgamloop}). Apart from the trivial cases
of the $2\to2$ processes (\ref{HH}) and (\ref{WH}), the
(numerical) integrations over the final state phase
space (and photon momentum fractions, see below)
have been performed by a variety of methods, for cross-checking
purposes: by using {\tt VEGAS} \cite{VEGAS}, {\tt RAMBO} 
\cite{RAMBO} and Metropolis \cite{Metropolis}. In the case of process
(\ref{HH}), we have found agreement with previous literature.

For the 2HDM parameters, we assumed the MSSM throughout.
For the SM ones, we adopted the following:
$m_b=4.25$ GeV, $m_t=175$ GeV, $m_e=0.511$ MeV, $m_\tau=1.78$ GeV,
$m_\nu=0$, $M_{W^<pm}=80.23$ GeV, $\Gamma_{W^\pm}=2.08$ GeV, 
$M_{Z^0}=91.19$ GeV,
$\Gamma_{Z^0}=2.50$ GeV, $\sin^2\theta_W=0.232$. The top quark width $\Gamma_t$
was evaluated at leading order for each value of $M_{H^\pm}$ and
$\tan\beta$. Neutral and charged Higgs masses and widths were calculated for 
given values of $M_{A^0}$ and $\tan\beta$ using the {\tt HDECAY}
package \cite{hdecay}, with the SUSY masses and and the Higgsino
parameter $\mu$ being set to 1 TeV, while the (universal) trilinear couplings
have been set to zero. (Hence, we only exploit here the MSSM mass
relations among the Higgs states, rather than investigating the effects
of new SUSY states.)

The back-scattered photon flux has been worked out 
in \cite{backscattering}, where all details of the derivation can be
found. For brevity,
we do not reproduce here those formulae, rather we simply 
recall to the un-familiar reader the basic features of
$\gamma\gamma$ scatterings initiated by laser light at $e^+e^-$
LCs. 
We assume that the laser back-scattering parameter $z$ of \cite{backscattering}
 assumes its maximum value,   $z\equiv z_{\rm{max}} = 2(1 +
\sqrt{2}) \simeq 4.828$. 
In fact, with increasing $z$ the high energy photon spectrum 
becomes more mono-chromatic. However,
for $z > z_{\rm{max}}$, the probability of $e^+e^-$ pair creation increases,
resulting in larger photon beam degradation.
The reflected photon beam carries off only a fraction $x$ of the 
electron/positron energy, with $x_{\rm{max}} = z/(1+z) \simeq 0.8$, while 
$x_{\rm{min}}= M_{\mathrm{X}}/\sqrt{s_{e^+e^-}}$, where
$M_{\mathrm{X}}$ is the rest mass in the final state of 
(\ref{HH})--(\ref{WH}).
Finally, one can cast the production cross sections
in the following form:
\begin{equation}
\sigma_{e^+e^-\to \gamma \gamma \to X} (s) 
= \int dx_+ dx_- F_+^{\gamma}(x_+) 
                   F_-^{\gamma}(x_-)
\hat\sigma_{\gamma \gamma \to X} (\hat s), 
\end{equation}
where $x_{+(-)}$ is the electron(positron) momentum fraction carried
by the emerging photon,
$ x_+x_- = {\hat{s}}_{\gamma\gamma}/s_{e^+e^-}$, 
with $s_{e^+e^-}$(${\hat s}_{\gamma\gamma})$ being the centre-of-mass (CM)
energy squared of the $e^+e^-$($\gamma\gamma $) system, 
and $F^\gamma_\pm(x_\pm)$
the photon distribution functions, defined in terms of
$ x_\pm$.
(As  $\gamma$-structure functions 
we have used those of Ref.~\cite{backscattering}.)

\clearpage

\clearpage

\clearpage

\begin{figure}[!t]
  \begin{center}
  \epsfig{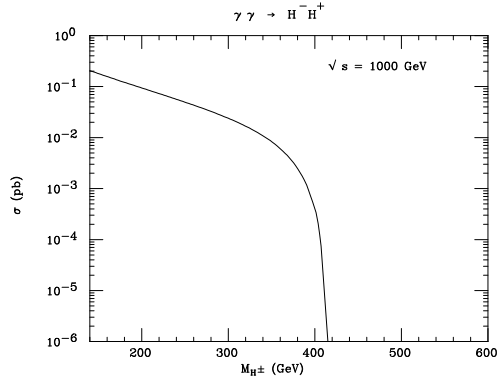}\\
  \caption{\small Total cross sections for process (\ref{HH})
      at $\sqrt s_{e^+e^-}=1$ TeV.}
  \label{cross_HH}
  \end{center}
  \vspace*{-0.5cm}
  \end{figure}

\begin{figure}[!t]
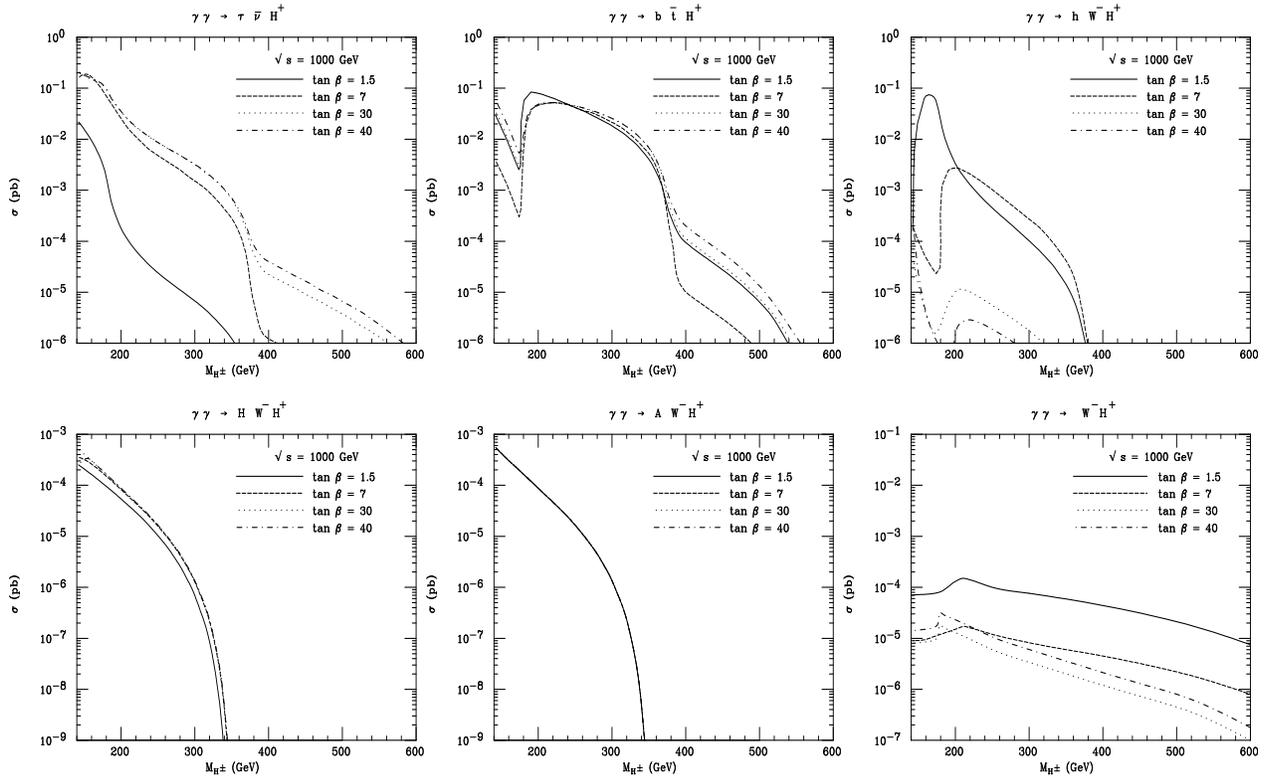

\begin{center}
\begin{minipage}[b]{.33333\linewidth}
\vspace{-0.75truecm}\centering\epsfig{file=gamgamtnhpm_1000.ps,angle=90,height=5cm,width=\linewidth}
\end{minipage}\hfil
\begin{minipage}[b]{.33333\linewidth}
\vspace{-0.75truecm}\centering\epsfig{file=gamgambthpm_1000.ps,angle=90,height=5cm,width=\linewidth}
\vskip-2.0cm
\end{minipage}\hfil
\begin{minipage}[b]{.33333\linewidth}
\vspace{-0.75truecm}\centering\epsfig{file=gamgamhpmwh02_1000.ps,angle=90,height=5cm,width=\linewidth}
\end{minipage}\hfil
\vspace*{1.0truecm}
\begin{minipage}[b]{.33333\linewidth}
\vspace{-0.75truecm}\centering\epsfig{file=gamgamhpmwh01_1000.ps,angle=90,height=5cm,width=\linewidth}
\end{minipage}\hfil
\begin{minipage}[b]{.33333\linewidth}
\vspace{-0.75truecm}\centering\epsfig{file=gamgamhpmwh03_1000.ps,angle=90,height=5cm,width=\linewidth}
\end{minipage}\hfil
\begin{minipage}[b]{.33333\linewidth}
\vspace{-0.75truecm}\centering\epsfig{file=gamgamwhpm_1000.ps,angle=90,height=5cm,width=\linewidth}
\end{minipage}\hfil
\caption{\small Total cross sections for (clockwise) processes 
(\ref{tnH}), (\ref{btH}), (\ref{H02WH}), (\ref{H01WH}), 
(\ref{H03WH}) [here, the four curves in the plot coincide within graphical
resolution] and (\ref{WH}) plus c.c. at $\sqrt s_{e^+e^-}=1$ TeV.}
\label{cross_H}
\vspace*{-0.5cm}
\end{center}
\end{figure}

The cross sections of processes (\ref{HH})--(\ref{WH})
can be found in Figs.~\ref{cross_HH}--\ref{cross_H},
respectively, for four reference choices of $\tan\beta$.
In all our plots and in the discussion, charge conjugated (c.c.)
contributions are always included. For brevity, we limit ourselves
to the representative case of $\sqrt s_{e^+e^-}=1$ TeV, noting
that the maximal energy achievable in $\gamma\gamma$ mode
is $\sqrt s_{\gamma\gamma}\approx 0.8~
\sqrt s_{e^+e^-}$. 

If one recalls the typical pattern of the charged Higgs bosons decays rates
into SM particles (see, e.g., \cite{BRs}), it is clear that some of
the final states considered in processes (\ref{tnH})--(\ref{WH}) can 
proceed via reaction (\ref{HH}) as intermediate state.
Hence, one should take care of avoiding double-counting
the $H^\pm$ production rates. For example,
in Fig.~\ref{cross_H} the $H^-\to\tau^-\bar\nu_\tau$ decay is always open
(top-left), the $H^-\to b \bar t$ one shows up at $M_{H^\pm}\approx m_t$ 
(top-middle) and
the $H^-\to h^0 W^-$ channel appears at small $\tan\beta$
when $M_{H^\pm}\lsim m_t$ (top-right).

Bearing this consideration in mind, one immediately realises
the dominance of $H^+H^-$ production whenever 
$\sqrt s_{\gamma\gamma}>2M_{H^\pm}$, independently of $\tan\beta$,
as expected. At and above the threshold point $\sqrt s_{\gamma\gamma}
\approx 2M_{H^\pm}$ (where pair production has extinguished),
there are three competing channels that can give sizable
signals: $\tau^-\bar\nu_\tau H^+$, $b\bar t H^+$ and $W^-H^+$,
the first two being largest at large $\tan\beta$ and the third
at small values of the latter\footnote{Some limited coverage at intermediate
values of $\tan\beta$ also exists via the one-loop mode (\ref{WH}), 
though with production cross sections that are one order of magnitude
smaller than at the lower end of the $\tan\beta$ interval.}. 
Final states of the type
$h^0 W^-H^+$ (for small to intermediate $\tan\beta$), 
$H^0 W^-H^+$ and
$A^0 W^-H^+$ (both also for large $\tan\beta$)
are only relevant for $M_{H^\pm}\lsim m_t$, where
they may well be useful in testing triple and quartic vertices involving
gauge and Higgs bosons. 

The total production rates in the region 
$\sqrt s_{\gamma\gamma}\lsim 2M_{H^\pm}$ are not very large, as they never 
exceed the fraction of femtobarns. 
After 1 ab$^{-1}$ of accumulated luminosity, one should 
expect at best 100 events or so, both at small and large $\tan\beta$.
Moreover, given the dependence upon this parameter of the
three leading modes, the intermediate $\tan\beta$ region 
(i.e., around 7 or so) would have little coverage, only through
charged Higgs production in association with a $W^-$ boson,
yielding typical production rates that are one order of magnitude
smaller than those seen for extreme values of this parameter (1.5
and 40).

Such small cross sections inevitably require one to select the dominant
decay channel of a heavy charged Higgs boson, i.e.,
$H^+\to t\bar b $ \cite{BRs}. Therefore, the leading signal signatures
would be
\begin{equation}\label{bbWtaunu}
b\bar b W^+ \tau^-\bar\nu_\tau, 
\end{equation}
\begin{equation}\label{bbbbWW}
b\bar b b\bar b W^+ W^-, 
\end{equation}
\begin{equation}\label{bbWW}
b\bar b W^+ W^-, 
\end{equation}
for processes (\ref{tnH}), (\ref{btH}) and (\ref{WH}), respectively.
In each case, one should expect the (irreducible) background to be
dominated by top-quark pair production and decay, i.e., 
\begin{equation}\label{tt}
\gamma\gamma\to t\bar t \to b\bar b W^+W-,
\end{equation}
possibly followed by
\begin{equation}\label{Wtaunu}
W^-\to \tau^-\bar\nu_\tau 
\end{equation}
and by gluon radiation as well, eventually yielding two additional $b$-quarks:
\begin{equation}\label{radiation}
t\bar t \to g^* b\bar b W^+W^- \to b\bar b b\bar b W^+W^-.
\end{equation}

Given the large hadronic activity associated with multiple production
of $b$-quarks, one would presumably require semi-leptonic decays
of the $W^+W^-$ and $W^+\tau^-\bar\nu_\tau$ systems, into electrons
and/or muons. In the case of the first signature, (\ref{bbWtaunu}),
it has been shown in Ref.~\cite{medetect} that, for the case
of $e^+e^-$ collisions, the signal extraction above the $t\bar t$
noise should be feasible
in a region of 50 to 100 GeV (depending on $\tan\beta$,
for values between 30 and 40) above
the threshold at $\sqrt s_{e^+e^-}\approx 2 M_{H^\pm}$, with
statistical significances between $3\sigma$ and $5\sigma$, in correspondence
of $1$ and $5$ ab$^{-1}$ of accumulated luminosity. Given that
the starting signal-to-background ratio ($S/B$) is here not
much different from the case of the corresponding $e^+e^-$ initiated
process (the signal here also being burdened by top-antitop production
and decay as dominant background), one 
should expect the same happening in the context
of photon-photon collisions, albeit with a reduced charged Higgs
mass scope, since the CM energy is smaller in this case (assuming
a contemporaneous running in $e^+e^-$ and $\gamma\gamma$ modes).

No explicit signal-to-background analysis exists yet for the
other two signatures in the context of electron-positron annihilations,
although there is some work in progress \cite{progress}. While one
may reasonably suppose that a selection strategy similar to
the one adopted in Ref.~\cite{medetect} would also work 
for the $\gamma\gamma\to W^- H^+$ process\footnote{If anything,
notice that the additional source of missing energy
due to the decays $\tau^-\to \pi^- X$ or $\tau^-\to \ell^- X$
($\ell=e,\mu$)
affecting the $\gamma\gamma\to\tau^-\bar\nu_\tau H^+$ process
is here largely absent.}, the same conclusion is not immediately evident
for the $\gamma\gamma\to b\bar t H^+$ channel. However, here one
could even improve on the results achievable in the case
of the signatures in (\ref{bbWtaunu}) and (\ref{bbWW}),
since the probability of a gluon splitting into $b\bar b$
pairs in (\ref{bbbbWW}) is rather small \cite{QCD}, even at the energies
at which the QCD gauge boson could be emitted in top pair production
and/or decay at future LCs. Indeed, $b\bar b b\bar b W^+W^-$
final states with semi-leptonic gauge boson decays have already been 
considered in the context of charged Higgs boson searches at the
LHC and proven to be accessible over the `pure' QCD background $q\bar q,gg\to
b\bar b t\bar t$ \cite{4b}.

Before closing, we comment on the energy dependence
of the three leading signal processes discussed above. In general,
it is the $\gamma\gamma\to b\bar t H^+$ channel that is the most
sensitive to the value of the CM energy, rather than the $\gamma\gamma\to
\tau^-\bar\nu_\tau H^+$ and $\gamma\gamma \to W^-H^+$ modes, because
of the large mass of the top quark (in comparison
to $m_\tau$ and $M_{W^\pm}$): the larger(smaller) $\sqrt s_{e^+e^-}$ the 
more(less) relevant process (\ref{btH}) becomes with respect to
reactions (\ref{tnH}) and (\ref{WH}).


In summary, total cross sections of heavy charged Higgs bosons
with mass similar to or larger than approximately half the
collider CM energy and produced via $\gamma\gamma$ modes compare
well to the corresponding $e^+e^-$ ones in most cases. 
In absolute terms, the latter are larger at smaller energies
whereas the former grows relatively with $\sqrt s_{e^+e^-}$, due to the
respective $s$- and $t,u$-channel dependence. When compared,
the two modes display a similar potential in accessing $H^\pm$ states 
with $2M_{H^\pm}\gsim \sqrt s_{e^+e^-/\gamma\gamma}$, the latter being
singly produced at a rate of 
${\cal O}(10^{-1}~{\rm{fb}})$ at best.
It will presumably be the interplay between the
typical mass scale of the charged Higgs bosons (that one could, e.g.,
either have a direct hint of from data or else estimate indirectly 
within the MSSM from the measured value of $M_{h^0}$
 at Tevatron and/or the LHC) and the machine performance
in producing mono-chromatic Compton back-scattered photons that
will eventually dictate whether to put
more effort in $\gamma\gamma$ or $e^+e^-$ analyses in the quest
for such particles at next generation LCs. 

However, the running time to 
be spent on each mode will most likely depend on the measured
value of $M_{h^0}$. On
the one hand, it should be recalled that in electron-positron annihilations
the CM energy is typically higher but 
the lightest Higgs boson is always produced in association with some
other particles (hence, with a phase space
suppression): a $Z^0$ (Higgs-strahlung), a $\nu_e\bar\nu_e/e^+e^-$
pair ($W^+W^/Z^0Z^0$ fusion) or the pseudoscalar Higgs state
(pair production). On the other hand, in photon-photon scatterings,
$h^0$ states are produced singly, via a loop of charged (s)particles, but with
a reduced energy and possibly, if the Higgs width is rather small, 
also off-resonance. Whichever the case, should the close investigation
of $h^0$ (and, possibly, $H^0$ and $A^0$) signatures need to be supported 
by the detection of charged Higgs states in order to clarify the nature
of the EW symmetry breaking, a LC with the option of photon
beams will be well placed in pursuing this task, over a considerable 
$M_{H^\pm}$ range, provided the value of $\tan\beta$ is either large or 
small.

\noindent\underbar{\sl Acknowledgements}~
The authors are grateful to Kosuke Odagiri for his many contributions
in the early stages of this project.
SM is indebted to the theory group at KEK for kind hospitality while part
of this work was being carried out.

 \end{document}